# Dynamic radiation force of a pulsed Gaussian beam acting on a Rayleigh dielectric sphere

Li-Gang Wang [1, 2, 3, *], and Cheng-Liang Zhao [1]

[1] *Department of Physics, Zhejiang University, Hangzhou, 310027, China*
[2] *Centre for Quantum Physics, COMSATS Institute of Information Technology, Islamabad, Pakistan*
[3] *Center of Optics, Department of Physics, The Chinese University of Hong Kong, Shatin, Hong Kong, China*
[*]*Corresponding author:* sxwlg@yahoo.com.cn

**Abstract:** We investigate the dynamic evolution of the radiation forces produced by the pulsed Gaussian beams acting on a Rayleigh dielectric sphere. We derive the analytical expressions for the scattering force and all components of the ponderomotive force induced by the pulsed Gaussian beams. Our analysis shows that the radiation force, for both the transverse and longitudinal components, can be greatly enhanced as the pulse duration decreases. It is further found that for the pulse with long pulse duration, it can be used for the stable trapping and manipulating the particle, while for the pulse with short pulse duration it may be used for guiding and moving the small dielectric particle. Finally we discuss the stability conditions of the effective manipulating the particle by the pulsed beam.

## 1. Introduction

In 1970, Askhkin [1] first demonstrated the optical trapping of particles using the radiation force produced by the focused continuous-wave (CW) Gaussian beam. Since then the optical trapping or tweezers has become a powerful tool for manipulating dielectric particles [2, 3], biological cells [4-7], and neutral atoms [8, 9].

Usually optical trapping or tweezers in many experiments are constructed by using the CW laser. It is well known that the CW laser, with the power of a few milliwatts, can only produce the radiation force with an order of a few *p*N to manipulate the micro-sized particles. Recently, Ambardekar et al. [10] and Deng et al. [11] used a pulsed laser to generate the large gradient force (up to 100*p*N) within a short duration (~45μs) for overcoming the adhesive interaction between the particles and the surface. Little et al. [12] and Agate et al. [13] have made a comparison between the femtosecond and the CW optical tweezers, and have pointed out that femtosecond optical tweezers are as effective as CW optical tweezers.

However, there is no any investigation about "how does the pulse duration affect on the radiation force acting on a Rayleigh particle?" The radiation forces as electromagnetic forces have been discussed in detail [14, 15]. In this paper, we have derived the analytical expressions for the longitudinal and transverse components of the radiation force under the Rayleigh approximation. Based on the derived formula, we analyze the effect of the pulse duration on the physical manipulating impact on the micro-sized particles. At last, we show the stability conditions for effective manipulating the particles.

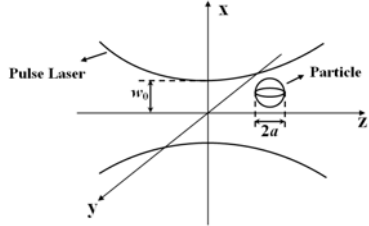

Fig. 1. Schematics of a pulse laser radiating on a particle.

## 2. Radiation force induced by a pulsed Gaussian beam

As shown in Fig. 1, a paraxial pulsed Gaussian beam is radiating on a dielectric sphere. The polarization direction of the pulsed electric field is assumed to be along the $x$ direction. The expression for the electric field of the paraxial pulsed Gaussian beam is given by

$$\vec{E}(\rho,z,t) = \hat{x} E(\rho,z,t)$$
$$= \hat{x} \frac{iE_0}{i+2z/kw_0^2} \exp[i\omega_0 t - ikz - \frac{i2kz\rho^2}{(kw_0^2)^2 + 4z^2} - \frac{(kw_0)^2 \rho^2}{(kw_0^2)^2 + 4z^2}] \exp[-\frac{(t-z/c)^2}{\tau^2}], \quad (1)$$

where $w_0$ is the beam waist at the plane $z=0$, $\rho$ is the radial coordinate, $\hat{x}$ is the unit vector of the polarization along the $x$ direction, $k = 2\pi/\lambda$ is the wave number, $\tau$ is the pulse duration, and $\omega_0$ is the carrier frequency. Eq. (1) is applicable under the narrow-spectral condition that the pulse spectrum width (proportional to $1/\tau$) is much smaller than $\omega_0$. For the fixed input energy $U$ of a single pulsed beam, the constant $E_0$ is determined by $E_0^2 = 4\sqrt{2}U/[n_2\varepsilon_0 cw_0^2(\pi)^{3/2}\tau]$, where $c = 1/(\varepsilon_0\mu_0)^{1/2}$ is the light speed in vacuum, $\varepsilon_0$ and $\mu_0$ are the dielectric constant and permeability in the vacuum, respectively, and $n_2$ is the refractive index of the surrounding medium. The corresponding magnetic field under the paraxial approximation can be given by

$$\vec{H}(\rho,z,t) \cong \hat{y} n_2 \varepsilon_0 c E(\rho,z,t), \qquad (2)$$

here we have neglected the $z$ component of the magnetic field under the paraxial approximation. The pulse intensity or irradiance is defined as a time-averaged Poynting vector:

$$\vec{I}(\rho,z,t) \equiv <\vec{S}(\rho,z,t)>_T = \hat{z}I(\rho,z,t) = \hat{z}\frac{P}{1+4\tilde{z}^2}\exp[-\frac{2\tilde{\rho}^2}{1+4\tilde{z}^2}]\exp[-2(\tilde{t}-\frac{\tilde{z}kw_0^2}{c\tau})^2], \qquad (3)$$

where $P = 2\sqrt{2}U/[(\pi)^{3/2}w_0^2\tau]$, and $(\tilde{\rho},\tilde{z},\tilde{t}) = (\rho/w_0, z/kw_0^2, t/\tau)$ are normalized temporal spatial coordinates, and $\hat{z}$ is the unit vector in the beam propagating direction.

For the sake of simplicity, we assume that the radius $a$ of the particle is much smaller than the wavelength of the pulse, i. e., $a \ll \lambda$. In this case, the Rayleigh approximation is applicable and the dielectric particle can be seen as a point dipole. Under this approximation, for the CW laser, the radiation force includes the scattering force $\vec{F}_{scat}$ and gradient force $\vec{F}_{grad}$, where $\vec{F}_{grad}$ arises from the inhomogeneous field distribution [14, 17]; for the pulse, $\vec{F}_{grad}$ is a part of the ponderomotive force. In a dilute medium (without dipole-dipole interaction), the ponderomotive force $\vec{F}_p$ is simply the Lorentz force [14, 16-17]:

$$\vec{F}_p(\rho,z,t) = [\vec{p}(\rho,z,t)\cdot\nabla]\vec{E}(\rho,z,t) + [\partial_t\vec{p}(\rho,z,t)]\times\vec{B}(\rho,z,t) = \vec{F}_{grad} + \vec{F}_t, \qquad (4)$$

where $\vec{p} = \beta\vec{E}$ is the dipole moment, $\beta = 4\pi n_2^2\varepsilon_0 a^3[(m^2-1)/(m^2+2)]$ is the polarizability of a spherical particle in the Rayleigh regime, and $m = n_1/n_2$ (here $n_1$ is the refractive index of the particle). Obviously, for the CW laser, the second term in Eq. (4) is always *zero*. In our case, for the short pulse, except for $\vec{F}_{grad}$, the second term $\vec{F}_t$ of Eq. (4) may also play an important role for manipulating the small particles. Substituting Eqs. (1) and (2) into Eq. (4), we can obtain all components of the ponderomotive force as follows:

$$\vec{F}_{grad,\rho} = -\hat{\rho}2\beta I(\rho,z,t)\tilde{\rho}/[cn_2\varepsilon_0 w_0(1+4\tilde{z}^2)], \qquad (5a)$$

$$\vec{F}_{grad,z} = -\hat{z}\frac{2\beta I(\rho,z,t)}{n_2\varepsilon_0 ckw_0^2}[\frac{\tilde{z}k^2w_0^4}{c^2\tau^2} - \frac{k\tilde{t}w_0^2}{c\tau} + \frac{2\tilde{z}(1+4\tilde{z}^2-2\tilde{\rho}^2)}{(1+4\tilde{z}^2)^2}], \qquad (5b)$$

$$\vec{F}_t = -\hat{z}8\mu_0\beta I(\rho,z,t)\tilde{t}/\tau + \hat{z}8\tilde{z}\mu_0\beta I(\rho,z,t)kw_0^2/(c\tau^2), \qquad (5c)$$

where $\hat{\rho}$ is the unit vector of the radial direction. For $\vec{F}_{scat}$, it is proportional to light intensity and is along the $+z$ direction [17],

$$\vec{F}_{scat}(\rho,z,t) = C_{pr}<\vec{S}(\rho,z,t)>_T/(c/n_2) = \hat{z}(n_2/c)C_{pr}I(\rho,z,t), \qquad (6)$$

where $C_{pr} = (8\pi/3)(ka)^4 a^2[(m^2-1)/(m^2+2)]^2$ is the cross section of the radiation pressure of the spherical particles in the Rayleigh regime. From Eqs. (5) and (6), we find that the magnitude of the radiation forces, especially for the longitudinal components $\vec{F}_{grad,z}$ and $\vec{F}_t$, is greatly affected by the pulse duration $\tau$. From Eqs. (5a) and (5b), both the transverse and longitudinal components of the gradient force act as the restoring forces pointed towards the pulse center for the particle with $m > 1$ although the magnitude of these forces will be changed with different $\tau$. In the following, we find that whether there is a stable trapping strongly depends on the duration $\tau$ because $\bar{F}_t$ is strongly affected by $\tau$. For smaller $\tau$, the second term of Eq. (5c) can be greatly increased in manner of $1/\tau^2$.

## 3. Numerical simulation and analysis

Without loss of generality, in all our simulations we choose the following parameters [18]: $\lambda = 0.514\,\mu m$, $m = n_1/n_2 = 1.592/1.332$ (for example, the small glass bead and water), $w_0 = 1\,\mu m$, $a = 5$ nm, and the input pulse power is fixed to be $U = 0.1\,\mu J$.

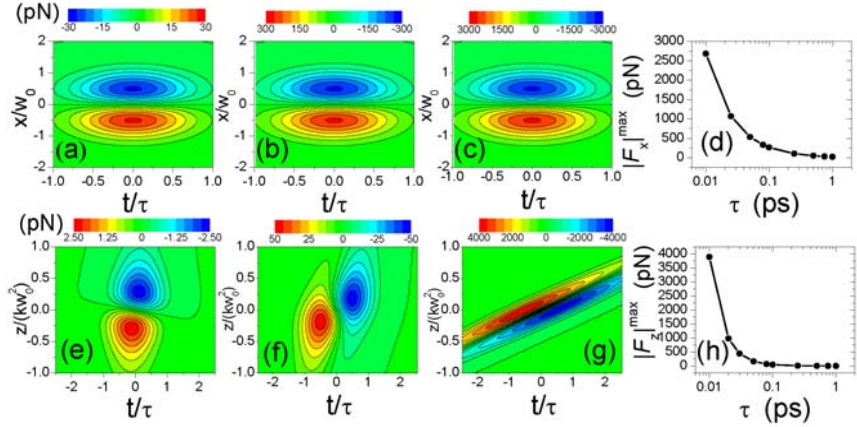

Fig. 2. The temporal evolutions of (a-c) the transverse and (e-f) the longitudinal radiation forces for the pulses with different durations: $\tau = 1$ ps for (a) and (e), $\tau = 0.1$ ps for (b) and (f), $\tau = 0.01$ ps for (c) and (g). (d) and (h) show the dependence of both the maximal transverse and longitudinal radiation forces on the duration $\tau$, respectively.

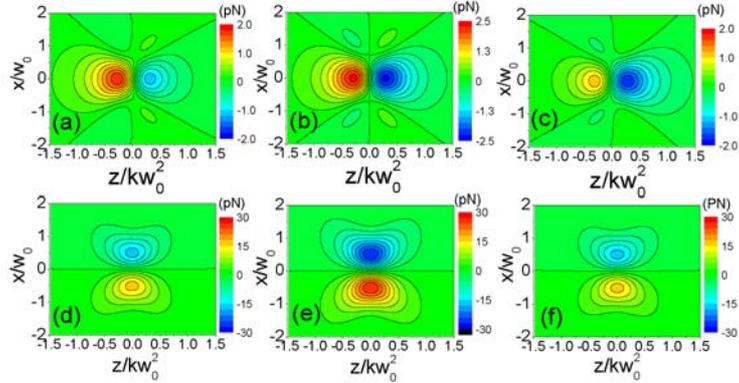

Fig. 3. The dynamic distributions of (a-c) the longitudinal and (d-e) the transverse radiation forces for the pulses with duration $\tau = 1$ ps at different times: $t = -0.5\tau$ for (a) and (d), $t = 0$ for (b) and (e), and $t = 0.5\tau$ for (c) and (f).

First, let us see the temporal evolution of the radiation force acting on the small particle. Figures 2(a), 2(b), and 2(c) plot the dynamic evolution of the transverse radiation force. It shows that there is a trapping region in the transverse direction, and the maximal transverse radiation force is enhanced as $\tau$ decreases [see Fig. 2(d)]. Figures 2(e), 2(f) and 2(g) plot the dynamic evolution of the longitudinal radiation force, $\bar{F}_z = \bar{F}_{scat} + \bar{F}_{grad,z} + \bar{F}_t$, i.e., all forces along the $z$ axis. As $\tau$ becomes small, the maximal $|\bar{F}_z|$ also increases greatly [see Fig. 2(h)], and the dynamic behavior of $\bar{F}_z$ changes dramatically from stably to unstably along the z axis. Therefore, for the pulse with large $\tau$, the force is determined by the gradient force, which forms a stable trapping region at the pulse center; in this situation, the gradient force is much larger than $\vec{F}_{scat}$ and $\vec{F}_t$, thus it is very similar to the case of the CW laser [2, 17-20]. For the pulse with small $\tau$, although the transverse radiation force provides the restoring force to pull the particle back to the stable point on the $z$ axis, the longitudinal radiation force will make the particle accelerate and decelerate along the $+z$ direction.

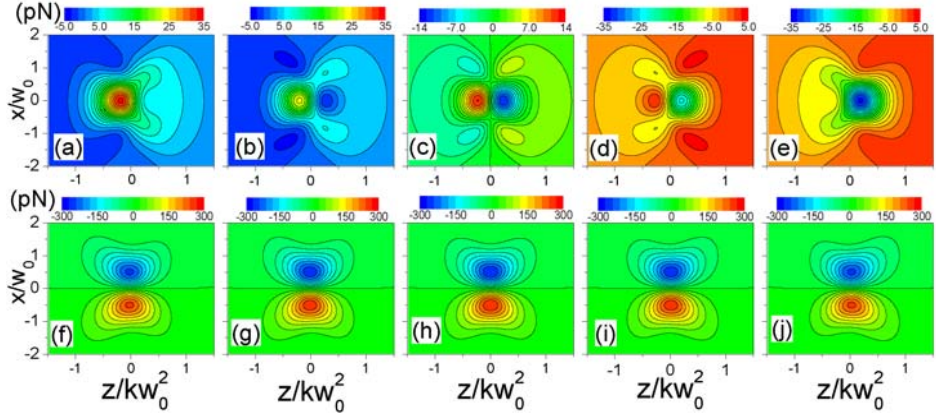

Fig. 4. The dynamic distribution of (a-e) the longitudinal and (f-j) transverse radiation forces for the pulse with $\tau = 0.1\,ps$ at different times: $t = -0.2\tau$ for (a) and (f), $t = -0.1\tau$ for (b) and (g), $t = 0$ for (c) and (h), $t = 0.1\tau$ for (d) and (i), and $t = 0.2\tau$ for (e) and (j).

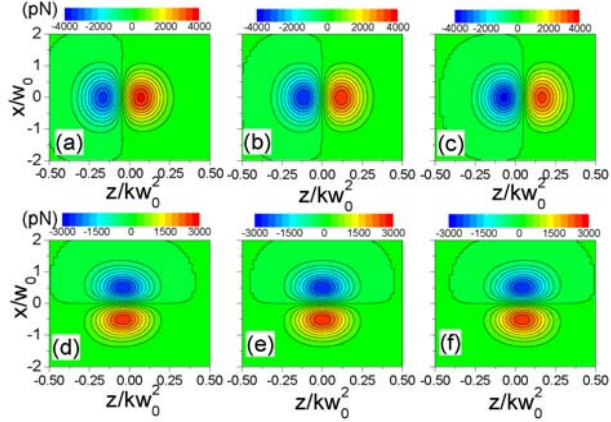

Fig. 5. The dynamic distribution of (a-c) the longitudinal and (d-f) transverse radiation forces for the pulse with $\tau = 0.01\,ps$ at different times: $t = -0.2\tau$ for (a) and (d), $t = 0$ for (b) and (e), and $t = 0.2\tau$ for (c) and (f).

Now we turn to discuss how the pulse duration affects on the distribution of the radiation force. Figure 3 shows the dynamic distributions of the transverse and the longitudinal radiation forces acting on the Rayleigh dielectric sphere for the pulse with $\tau = 1\,ps$ at different times. It is clear that for the pulse with larger $\tau$, the radiation force has a stable trapping region, which is located near the focused center of the pulse and the trapping is enhanced as the pulse arrives at the focused center $z = 0$. Therefore in this case the pulse may trap and manipulate the micro-sized particles as effectively as the CW laser [12, 13]. With the decreasing of the pulse duration, see Fig. 4, the stable trapping region becomes small and the trapping time becomes short, although the transverse force still provides the restoring force in the transverse direction. For the pulse with very short duration, from Fig. 5, although the pulse cannot be used as an optical trap to manipulate the small particle, it may be used to accelerate the particle because (i) during the first half pulse, the particle will be confined around the $z$ axis and be accelerated along the $+z$ axis, and (ii) during the other half pulse, although the particle will be decelerate it still moves along the $+z$ axis. It also should be noticed that the magnitude of the radiation force are greatly enhanced for the short pulse (also see Fig. 2). In this sense, the pulsed beam may be used to accelerate and levitate the particle for overcoming the van der Walls force from the surface of the vessel [10, 11].

**4. Conditions for the stable manipulation**

In the above discussion, our analysis shows that the radiation force of the pulsed beam may be used to manipulate the Rayleigh dielectric sphere. In order to know whether the particle could be effectively controlled by the pulse, first we estimate the potential well induced by the radiation force. The potential well must be deep enough to overcome the kinetic energy of the particle due to thermal fluctuation. The condition can be given by using the Boltzmann factor [2, 8]: $R_{thermal} = \exp(-U_{max}/k_B T) << 1$, where $U_{max} = \pi \varepsilon_0 n_1^2 a^3 |(m^2-1)/(m^2+2)| \cdot |\vec{E}|_{max}^2$ is the maximum of the potential well, $k_B$ is the Boltzmann constant, and $T$ is the absolute temperature of the ambient. In the above numerical examples, for the pulse with $\tau = 1\,ps$, we obtain $R_{thermal} \approx 0.68$ in the room temperature of 300 K. As $\tau$ decreases, $R_{thermal}$ becomes smaller and smaller, e. g., $R_{thermal} \approx 0.02$ for $\tau = 0.1\,ps$, and $R_{thermal} \approx 3.9 \times 10^{-17}$ for $\tau = 0.01\,ps$. Therefore the Brownian motion could be overcome or ignored and the particles can be manipulated by the radiation force of the pulsed beam with short duration.

Second, we consider the effect of the random diffusion of the small particle in the surrounding medium. As we know that during the interval of two pulses, the radiation force vanishes, and the particle moves under the gravity and the Brownian force. If the particle doesn't move out the trapping region during the pulse interval, then the radiation force of the second pulse can drag it back to the stable trapping region or manipulate the particle again. In our case, the density of the small glass bead is assumed to be $2.4 \times 10^3$ kg/m$^3$, then the gravity of the particle is about $1.2 \times 10^{-20}$ N, which is much smaller than the radiation force, therefore the effect of the gravity can be ignored for the small particle with $a = 5$ nm. When a particle moves freely inside the ambient, it suffers the Brownian motion. From the Stokes-Einstein relation, the diffusion coefficient is $D = k_B T/(6\pi\eta a)$, where the viscosity for water is $\eta = 7.977 \times 10^{-4}\,Pa \cdot s$ at $T = 300$ K. Therefore we obtain $D \approx 5.5 \times 10^{-11}$ m$^2$/s for $a = 5$ nm. If the pulse repetition rate is larger than 100Hz, the particle diffuses into an area about $0.55 \times 10^{-12}$ m$^2$ during the pulse interval. This diffusion area is smaller than the acting range of the radiation force. Therefore the particle could be effectively confined within the manipulating region of the pulsed beam.

**4. Conclusion**

In summary, we have theoretically investigated the dynamic radiation force produced by the pulsed Gaussian beam acting on a Rayleigh dielectric particle. We find that the radiation force of the pulse can increase greatly as the pulse duration decreases. For the pulsed beam with large pulse duration, it can be used for stable trapping and manipulating the particle; for the pulsed beam with short pulse duration it may be used for guiding and moving the small dielectric particle. Finally the conditions for effective trapping and manipulating the particle have been analyzed.

**Acknowledgments**

This work was supported by National Natural Science Foundation of China (No. 10604047), and Scientific Research Foundation of Zhejiang Province (G80611 and G20630).